\documentclass[12pt,preprint,dvips]{aastex}

\newcommand{\Teff}{$T_{\rm eff}$}

\shorttitle{Error Analysis for geometric BW method}
\shortauthors{M. Marengo et al.}

\begin{document}


\title{An Error Analysis of the Geometric Baade-Wesselink Method}

\author{Massimo Marengo, Margarita Karovska, Dimitar D. Sasselov}
\affil{Harvard-Smithsonian Center for Astrophysics, 60 Garden St.,
       Cambridge, MA 02138}
\email{mmarengo@cfa.harvard.edu, mkarovska@head-cfa.harvard.edu,
       sasselov@cfa.harvard.edu} 

\author{Mayly Sanchez}
\affil{Department of Physics, Harvard University, 17 Oxford St.,
       Cambridge, MA 02138}
\email{msanchez@physics.harvard.edu}

\begin{abstract}

We derive an analytic solution for the minimization problem in
the geometric Baade-Wesselink method. This solution allows deriving
the distance and mean radius of a pulsating star by fitting its
velocity curve and angular diameter measured interferometrically. The
method also provide analytic solutions for the confidence levels of
the best fit parameters, and accurate error estimates for the
Baade-Wesselink solution. Special care is taken in the analysis of
the various error sources in the final solution, among which the
uncertainties due to the projection factor, the limb darkening and the
velocity curve. We also discuss the importance of the
phase shift between the stellar lightcurve and the velocity curve as a
potential error source in the geometric Baade-Wesselink method. 
We finally discuss the case of the Classical
Cepheid $\zeta$~Gem, applying our method to the measurements derived
with the Palomar Testbed Interferometer. We show how a careful
treatment of the measurement errors can be potentially used to
discriminate between different models of limb darkening using 
interferometric techniques.

\end{abstract}




\section{Introduction}\label{sec-intro}

Since its conception, the Baade-Wesselink (BW) method
\citep{baade1926, wesselink1946} has been adopted as a
preferred way to measure the distance of pulsating stars. In its
classical formulation, the distance modulus of a star is derived
from the stellar physical radius, obtained by
integrating the stellar radial velocity curve, and its effective
temperature. The method has been applied to calibrate the
Period-Luminosity relation for different classes of pulsators, among
which RR Lyr\ae{} \citep{mcdonald1977, jones1992, cacciari1992,
fernley1994, bono1994}, Cepheids (see review by
\citealt{gautschy1987}), Delta Scuti and SX Phoenix variables 
\citep{meylan1986}. 

The main limitation of this technique, however, consists in finding a
reliable observable yielding an accurate estimate of the stellar
\Teff{} \citep{boehm-vitense1989}.
As a result, many variants of the BW method have been developed
to circumvent this problem, by using different combinations of colors
and band-passes (see e.g. \citealt{gieren1993, laney1995, ripepi1996,
balog1997}). Despite these ongoing observational efforts,
discrepancies in the BW distances derived with different colors are
still present \citep{laney1995, krockenberger1997}.

Recent progress in interferometric techniques have made
possible direct determination of angular diameter variations in
nearby pulsating stars. This achievement has resulted in the
development of a \emph{geometric} version of the Baade-Wesselink
method, which is in principle free of the limitations of the
color-based techniques. Applied to the galactic
Cepheids, this method should eventually lead to a reliable zero
point of the Cepheid distance scale \citep{sasselov1994}.

Attempts to measure the angular radial displacements of pulsating
Cepheids were made using the IOTA interferometer \citep{kervella2001},
the Palomar Testbed Interferometer (PTI, \citealt{lane2000, lane2002})
and the Naval Prototype Optical Interferometer (NPOI). The results
of these observations show that indeed the BW method can be used to
yield distances of nearby Cepheids with a much better accuracy than
other methods, including geometric parallax \citep{feast1997,
esa1997}. 

The geometric BW method is limited in its accuracy by uncertainties
due to observational errors and model dependencies. These model
dependencies derive from the need to translate the observed
visibilities into angular diameters, which require accurate
center-to-limb brightness profiles specific for the pulsating
star. These profiles can be derived from hydrodynamic models of the
stellar atmosphere (\citealt{marengo2002}, hereafter Paper~I). 
The  conversion of radial velocities into
radial displacement requires the knowledge of a projection factor
which is computed by modeling line formation in the pulsating
atmosphere \citep{sabbey1995}. To these model-dependent uncertainties
one should add the intrinsic errors of the interferometric
measurements, and possible irregularities in the periodic pulsational
behavior of the star. 

In this paper we revisit the geometric BW method, in order to discuss
the relative importance of various uncertainties, assessing their
contribution to the accuracy of the final distance determination.
We first derive an
original analytical solution for the BW method fitting procedure, which
simplify the determination of the best fit distance and average radius
from the observational data, and allows a more transparent analysis of
the individual error sources. We then apply our revised method on
$\zeta$~Gem PTI data \citep{lane2002}, and we conclude with a discussion
of the importance of the errors due to the
pulsational model uncertainties (projection factor and limb darkening)
and the phase shift between the lightcurve of the pulsator and
the dynamical phases of the radial velocities.


\section{Geometric Baade-Wesselink method}\label{sec-method}

The \emph{geometric} BW method allows deriving the distance and the
mean radius of a pulsating star by fitting the variations of its angular
diameter measured interferometrically, with
the radial displacement $\Delta R(\phi)$ of the stellar
photosphere. The radial displacement is in turn derived by integrating
the pulsational velocity over time:

\begin{equation}\label{eq-dR}
\Delta R(\phi) = \int^\phi_{\phi_0} p(\phi') \left[ v_r(\phi') -
\gamma \right] \hbox{d}\phi'
\end{equation}

\noindent
where $v_r(\phi)$ is the radial velocity, which should be corrected 
for the systemic velocity $\gamma$ and the projection factor $p(\phi)$
to yield the pulsational velocity. 

As explained in Paper~I the systemic velocity is
calculated by requiring the conservation of the radius over one
period. It takes into account the radial motion of the star relative
to the Solar System, and physical inconsistencies in the measurement
of the radial velocities with spectral data. An appropriate $p$-factor
is computed by modeling spectral line formation for a given pulsating
star, including spectral line asymmetries and the dynamics 
of the pulsation. Models for several classical Cepheids have been
computed by \citet{sabbey1995}, from which the pulsational
phase-dependent $p$-factors are derived. The variations of the radial
velocity with the pulsational phase are instead measured
spectroscopically. A recent compilation of radial velocity data for 40
Cepheids have been published by \citet{bersier1994b}, by using the
CORAVEL spectrophotometer. The measurements are provided as a Fourier
expansion of $v_r$, in terms of the pulsational phase $\phi$. The
integration of the $v_r(\phi)$ curve provides the required radial
displacement $\Delta R(\phi)$, used in combination with the
measured angular diameters in the geometric BW method.

The BW distance $D$ and average radius $R_0$ of the pulsating star are
solved as a two parameter $\chi^2$ minimization:

\begin{equation}\label{eq-chi2}
\chi^2(R_0,D) = \sum_i \left[ \frac{\Theta_i - \theta_i}{\sigma_i}
\right]^2 = \sum_i \left[ \frac{ 2 \frac{R_0 + \Delta R_i}{D} -
\theta_i}{\sigma_i} \right]^2 
\end{equation}

\noindent
where the index $i$ runs on the data points, $\theta_i$ is the angular
diameter measured at a certain phase $\phi_i$ and $\Delta R_i = \Delta
R(\phi_i)$ is the radial displacement derived with equation~\ref{eq-dR}
for the same
phase. The terms $\sigma_i$ are the errors associated with each
data point, including the uncertainty in the radial displacement.
Note that when converting the time of each individual data point
$\theta_i$ into the pulsational phase $\phi_i$ of the radial
displacement curve, special care is required to take into account
the phase shift observed between the stellar lightcurve and
the radial velocity curve from which $R(\phi_i)$ is derived. Note also
that the diameters $\theta_i$, when measured with an interferometer,
should consider the Limb Darkening (LD) of the stellar
atmosphere. As in the case of the $p$-factor, this quantity is also
affected by the stellar pulsation, and depends on the pulsational
phase. In paper~I a method was presented to compute wavelength
and phase dependent LD, using the same hydrodynamic models from which
the \citet{sabbey1995} p-factor was derived.

The $\chi^2$ fit for the geometric BW method is solved by minimizing
equation~\ref{eq-chi2} with respect to the average radius $R_0$ and
the distance $D$. The minimum $\chi^2$ is found at the stationary
point of $\chi^2(R_0,D)$:

\begin{equation}\label{eq-chi2der}
\left\{ \begin{array}{lll}
\partial_{R_0} \chi^2 & = & 0\\
\partial_D \chi^2 & = & 0
\end{array} \right.
\end{equation}

\noindent
which gives a system of two linear equations in $D$, quadratic in $R_0$:

\begin{equation}\label{eq-chi2sys}
\left\{ \begin{array}{lll}
D & = & \frac{2A}{B} R_0 + \frac{2C}{B}\\
D & = & 2 \frac{A R_0^2 + 2 C R_0 + F}{B R_0 + E}
\end{array} \right.
\end{equation}

\noindent
where the following coefficients are defined as:

\begin{equation}\label{eq-coeff}
\left\{ \begin{array}{lll}
A & = & \sum_i \left( \frac{1}{\sigma_i^2} \right)\\
B & = & \sum_i \left( \frac{\theta_i}{\sigma_i^2} \right)\\
C & = & \sum_i \left( \frac{\Delta R_i}{\sigma_i^2} \right)\\
E & = & \sum_i \left( \frac{\Delta R_i \theta_i}{\sigma_i^2} \right)\\
F & = & \sum_i \left( \frac{\Delta R_i^2}{\sigma_i^2} \right)\\
G & = & \sum_i \left( \frac{\theta_i^2}{\sigma_i^2} \right)
\end{array} \right.
\end{equation}

The system in equation~\ref{eq-chi2sys} describes two curves
intersecting at the minimum $\chi^2$. The angular coefficient of the first
curve, which is a first order polynomial, is $2A/B$. The second curve,
when linearized to the first order, has an angular coefficient also
very close to $2A/B$. This means that the two curves described by the
system in equation~\ref{eq-chi2sys} are almost parallel. Their
intersection will thus result in a large uncertainty along the $D/R_0
= 2A/B$ direction, and a much smaller error on the orthogonal
direction. 

The system in equation~\ref{eq-chi2sys} can be solved by direct
substitution, obtaining first an explicit solution for $R_0$ in terms
of the coefficients defined in equation~\ref{eq-coeff}. When the
solution for $R_0$ is substituted back into the second equation, the
quadratic terms cancel, leaving a unique solution for $D$. Thus the
best fit distance $\bar D$ and mean radius $\bar R_0$ are: 

\begin{equation}\label{eq-chi2sol}
\left\{ \begin{array}{lll}
\bar R_0 & = & \frac{C E - B F}{B C - A E}\\
\bar D & = & 2 \frac{B C^2 - A B F}{B^2 C - A B E}
\end{array} \right.
\end{equation}

The value of the $\chi^2$ for the best fit parameters is obtained by
substituting $\bar R_0$ and $\bar D$ into equation~\ref{eq-chi2},
obtaining the following expression:

\begin{equation}\label{eq-chi2min}
\chi^2_{min} = \left. \chi^2(R_0,D) \right|^{R_0 = \bar R_0}_{D = \bar D} = 
\left[ 4 A \left( \frac{R_0}{D} \right)^2 + 8 C \frac{R_0}{D^2}
+ 4 \frac{F}{D^2} - 4 B \frac{R_0}{D} - 4 \frac{E}{D} + G
\right]^{R_0 = \bar R_0}_{D = \bar D}
\end{equation}

The error region for the best fit solution is determined by the
confidence levels of the $\chi^2(R_0,D)$. Assuming a gaussian error
distribution, the 68\% confidence level (giving the 1$\sigma$ error)
is defined by the equation: 

\begin{equation}\label{eq-cont}
\chi^2(R_0,D) = \chi^2_{min} + \chi^2_\textrm{68\%}
\end{equation}

\noindent
where $\chi^2_\textrm{68\%} = 2.33$ for a two parameter fit (see
e.g. \citealt{frodesen1979}). The explicit form of the error contour
in terms of $R_0$ and $D$ is obtained from equation~\ref{eq-chi2min}
by substitution of $\chi^2_{min}$ with $\chi^2_{min} +
\chi^2_\textrm{68\%}$. Note that the equation describes a tilted
ellipse having the major axis oriented along $D = R_0$. 

Error bars for the $R_0$ and $D$ best fit solution, at a certain
confidence level, can be obtained analytically by determining the
intersection of the error ellipse with appropriate lines parallel to
the $R_0$ and $D$ axis. Each line intersects the ellipse in two points
(given by a quadratic equation). The intersections which provide the
error bars are the ones where these two points coincide; there are two
such intersections for each axis, each of them given by the solution
of a quadratic equation in $D$ and $R_0$ respectively. These solutions
can be solved explicitly in terms of the coefficients defined in
equation~\ref{eq-coeff} as:

\begin{equation}\label{eq-fitXerr}
X^\pm = \pm \bar X \pm \frac{\pm \beta_X - \sqrt{\beta_X^2 -
\alpha_X \gamma_X}}{\alpha_X}\\
\end{equation}

\noindent
where $X$ is either $D$ or $R_0$, and where the numeric coefficients
in the two cases are:

\begin{equation}\label{eq-Derrc}
\left\{ \begin{array}{lll}
\alpha_D & = & B^2 - A G + A \left( \chi^2 + \chi^2_\textrm{68\%}
\right)\\
\beta_D & = & 2 \left( A E - C B \right)\\
\gamma_D & = & 4 \left( C^2 - A F \right)
\end{array} \right.
\end{equation}

\noindent
and

\begin{equation}\label{eq-Rerrc}
\left\{ \begin{array}{lll}
\alpha_{R_0} & = & B^2 - A G + A \left( \chi^2 + \chi^2_\textrm{68\%}
\right)\\
\beta_{R_0} & = & B E + C \left( \chi^2 + \chi^2_\textrm{68\%} \right)
- C G\\
\gamma_{R_0} & = & E^2 + F \left( \chi^2 + \chi^2_\textrm{68\%}
\right) - F G
\end{array} \right.
\end{equation}

The best fit angular diameter is simply obtained as $\bar \Theta_0 =
\bar R_0 / \bar D$. Note that the expression for $\Theta = R_0 / D$ is
the equation for the major axis of the error ellipse. Lines parallel
to the ellipse major axis are the regions of constant angular size for the
generic $R_0$ and $D$ solutions; the error bars for $\bar \Theta_0$
are thus given by the intersections of these lines with the error
ellipse. The solution can then be found analytically, as in the case of
the error bars of $R_0$ and $D$, with the following coefficients:

\begin{equation}\label{eq-THerrc}
\left\{ \begin{array}{lll}
\alpha_{\Theta_0} & = & C^2 - A F\\
\beta_{\Theta_0} & = & B F - C E\\
\gamma_{\Theta_0} & = & E^2 + F \left( \chi^2 + \chi^2_\textrm{68\%}
\right) - F G
\end{array} \right.
\end{equation}


\section{Application to the Classical Cepheid
$\zeta$~Gem}\label{sec-discuss} 

To illustrate the procedure described in the previous section, and
analyze the relative importance of the various error sources in the
final solution, we can consider the case of the Classical Cepheid
$\zeta$~Gem, recently observed with the PTI interferometer 
\citep{lane2002}. 

The Uniform Disk (UD) angular diameters $\theta^{(UD)}_i$ obtained for
$\zeta$~Gem in the near-IR H-band using the PTI have been published in
Table~4 of \citet{lane2002}. Each diameter results from a number of
individual measurements (\emph{scans}) performed at a certain date (in
Julian Days, JD).
The JD of each measurement is converted into the
$\zeta$~Gem \emph{lightcurve} phase $\phi^{(i)}_L$ by using the
published period $P = 10.150079$ days and the zero-phase epoch $T_0 =
2,444,932.736$ \citep{lane2002}. As mentioned previously,
the fitting procedure requires reconciling the ``\emph{dynamical}'' phase
$\phi_V$ of the radial velocity curve with the lightcurve phase
$\phi_L$ of the observations. This is done by correcting for the
\emph{phase shift} $\Delta \phi = \phi_L - \phi_V$ observed between
the maximum luminosity and the minimum radius. From the data collected
by \citet{bersier1994a, bersier1994b} we derive $\Delta \phi \simeq
0.28$. This is the value we will use in the following discussion. 
Limitations of this approach for deriving the phase shift are
discussed in section~{ssec-shift}.

To derive an accurate distance with the geometric BW method, the
angular diameter measurements $\theta_i$ should be corrected for
the LD. For clarity, however, we first solve the BW fit using the
published UD diameter, deferring the discussion of the effect of the
LD, and the errors associated with it, to a separate section. For the
same reason, we adopt the value of the $p$-factor of 1.43 used in
\citet{lane2002}, which is the average value derived by
\citet{sabbey1995} over the $\zeta$~Gem pulsational cycle, consistent
with the model we adopted for computing the LD.

As discussed before, the terms $\sigma_i$ are the errors associated
with each data point. They are the geometric sum of the individual
errors for each observation:

\begin{equation}\label{eq-sigma}
\sigma_i^2 = \left[ \sigma_i^{(\theta)} \right]^2 + \left[
\sigma_i^{(\Delta R/D)} \right]^2
\end{equation}

\noindent
where $\sigma_i^{(\theta)}$ is the error of the interferometric
measurement and $\sigma_i^{(\Delta R/D)}$ is the error related to the
diameter variation. According to \citet{lane2002} the errors
in the interferometric PTI UD data vary between 0.01 and 0.06~mas
(10--60\% of the predicted amplitude of $\zeta$~Gem pulsations, of
$\sim 0.1$~mas),
depending on the quality of each measurement. Note that these errors
refer only to the determination of the UD diameters. When computing
the \emph{true} distance of the star using the diameters corrected for
the LD, we will also take into account the errors involved into the LD
correction. 

An estimate of the error related to $\Delta R/D$ is more difficult to
obtain. The first component in this error is due to the uncertainty in
the radial velocity measurement. From \citet{bersier1994a} we get that
the average error in the $\zeta$~Gem velocity measurement is about
0.4~km/s over velocities up to $\sim 20$~km/s, which takes into
account possible irregularities in the pulsational behavior of the
star. This is a relative
error of up to $\sim 2$\% (when the velocity is maximum), which
translates to a maximum uncertainty in the expected stellar
radius variation of $\sim 0.002$~mas for the predicted amplitude of
the $\sim 0.1$~mas $\zeta$~Gem pulsation.
Another contribution to this error is the
uncertainty in the $p$-factor; we use the same value of $1.43 \pm
0.06$ as \citet{lane2002}, which takes into account the variation
of \citet{sabbey1995}  $p(\phi)$ with the pulsational phase.
This adds a further $\sim 4$\% relative uncertainty on the measured
pulsation amplitude (equivalent to $\sim 0.004$~mas). The total error 
associated to the $\zeta$~Gem angular diameter variation due to
the uncertainties in the $\Delta R/D$ fitting function (again assuming
that the two error sources are not correlated, and add as their
geometric sum) is thus of $\sim 4.5$\%. This is equivalent to a
maximum error of $\sim 0.005$~mas for the predicted amplitude of the
$\zeta$~Gem pulsations. Note that this error, in general, is
smaller than the error of the current interferometric
data, but may become important in the future, given the promising
advances in the interferometric techniques. To include the
$\sigma_i^{(\Delta R/D)}$ in the BW fit, we have first solved
equation~\ref{eq-chi2sol} in order to have a preliminary estimate of
$D$, and then evaluated the $\sigma_i^{(\Delta R/D)} \simeq (0.06 \cdot
\Delta R_i / D)$ for each point. Finally, we have repeated the $\chi^2$
fitting procedure considering both error sources. 

The best fit results are summarized in Figure~\ref{fig1}, which shows
the BW distance and average radius for the PTI UD data,
and the error regions for the 90\% (external oval) and 68\% (inner
oval) confidence level. As expected, the error regions are ellipses
with the major axis oriented along the $R_0/D = const$ direction. The
ellipses are very narrow, resulting in large errors, individually, in
the best fit $R_0$ and $D$, but a small error in their ratio $\Theta_0
= R_0/D$. This is not surprising, since the accuracy of the
interferometric measurements is good enough to produce a reasonable fit
to the average angular radius of the pulsating star; a much higher
accuracy is instead required to measure the radial displacement, from
which $R_0$ and $D$ are measured with the BW method.
Equation~\ref{eq-chi2} can in fact be written as: 

\begin{equation}\label{eq-chi2-th}
\chi^2 = \sum_i \left[ \frac{\left( \Theta_0 + \frac{2 \Delta R_i}{D}
\right) - \theta_i}{\sigma_i} \right]^2
\end{equation}

Given that the maximum angular diameter variations are less than 10\%
of the average angular diameter, then for most phases $\Theta_0 \gg 2
\Delta R_i / D$. This means that the fit is essentially a
one-parameter fit for $\Theta_0$, which can be determined with good
accuracy. The measurement errors plays a much larger effect in
limiting the accuracy of $R_0$ and $D$ along $R_0/D = \Theta_0$.


\subsection{Phase Shift Determination}\label{ssec-shift}

Two basic questions remain open when discussing the accuracy of the
fit using the BW method. The first concerns the assumption that the
phases of the interferometric data and radial velocities are well
synchronized, e.g. that there is not an unknown phase shift component
between the luminosity lightcurve and radial pulsations. The second
question concerns the LD correction, which we need to apply to the UD
measurement to derive the correct BW distance.

As mentioned before, the BW method requires defining a common phase
reference for diameter observations and radial velocity data. This
is usually provided by the lightcurve phases $\phi_L$, with the zero-phase
set to coincide with the maximum luminosity. A complication can
however arise when observable quantities are related to phase
dependent parameters which are derived from dynamic modeling of pulsating
stars. Typical parameters are the $p$-factor \citep{sabbey1995} and
the LD correction (Paper~I). The traditional choice of the
lightcurve phase is not convenient when dealing with
time-dependent hydrodynamic models, e.g. the ones from which the
$p$-factor and the LD are computed. The emergent flux in a given
waveband is in fact a derived quantity (which can only be computed
with extensive radiative transfer modeling), and is usually not
available in hydrodynamic numeric computations.  A better approach
in this case is to choose the zero-phase reference based on
hydrodynamic quantities directly tied with observables, as the radial
velocity $v_r$. This leads to the ``dynamical phase'' $\phi_V$
mentioned before. 

The complication in reconciling the lightcurve and dynamical phase
references is due to the phase shifts occurring between the zero phases
in the two systems. This quantity can be obtained by carefully
measuring the lightcurve and velocity curve of the pulsating star, as
done by \citet{bersier1994a, bersier1994b}. This procedure yields for
$\zeta$~Gem the quantity $\Delta \phi \simeq 0.28$ used in the
previous section; different values should be expected for different
stars, since the phase shift is determined by the unique dynamics of
the pulsator. A different approach was followed by \citet{lane2002},
in which the phase shift was derived as a separate quantity in the BW
fit, on the assumption that having a third free parameter would produce
a better agreement between the measured angular diameters and the
radial displacement curve.

To assess the effects on the fit results by an independent
determination of the phase shift, we repeated the BW fit of the
$\zeta$~Gem PTI UD diameters leaving this parameter free. The results
are showed in Figure~\ref{fig2}, where the best fit is plotted in the
case of a free phase shift (dashed line), and a fixed phase shift (solid
line) as determined from \citet{bersier1994a, bersier1994b} data. The
best fit results with their errors are shown in Table~1.

The results show that the small difference in the phase
shift, although ``visually'' improving the fit of the data points, does
not change significantly the best fit results. The difference in $R_0$,
$D$ and $\Theta_0$ is well below the error bars resulting from the
fit, which suggests that the shift itself can be a statistical
deviation due to the uncertainties of the data points. It is however
interesting to investigate the possible causes of this discrepancy,
since an unaccounted phase shift would have profound implications in
terms of the evolution of pulsating stars.

Since the actual phase of interferometric data acquired at certain JD
is determined by folding many pulsational periods starting from the
zero phase epoch, an uncertainty in the period determination can play
a significant role in changing the measured phase shift. The typical
uncertainty for the Cepheid periods tabulated in \citet{bersier1994a}
is of the order of $\sim 10^{-4}$ days. The zero phase epoch
for $\zeta$~Gem is around day 2,444,933, while the observations were
made between JD 2,451,605 and JD 2,451,896. Clearly the 300 days
span in the time during which the observations were collected is not
important since it corresponds to less than 30 periods. The time lag
between the zero epoch and the data acquisition is instead of $\sim
7000$~days, which is approximately 690 periods. Given the accuracy of the
$\zeta$~Gem period, the possible shift is of the order of $\sim
0.07$~days. This is less than a 0.01 phase shift error, which is one
order of magnitude less than the unaccounted phase shift derived by
fitting the PTI data. Therefore, if the measured phase shift is
real, it is not explained by the current uncertainty in the
pulsational period. 

\citet{berdnikov2000} analyzed archival data of the maximum
luminosity epochs for a number of Classical Cepheids, in search for
secular variations of their period due to evolutionary changes. In the
case of $\zeta$~Gem, they found a decrease of 0.07~days over a time of
about 5000 pulsational cycles. This means a decrease of the period of
$1.4 \cdot 10^{-5}$~days per period, which is much less than the
period uncertainty adopted in this paper and in \citet{lane2002}.

Another unaccounted source of possible variations in the phase shift
involves the possibility that the interferometric angular diameter
may be slightly different (and off-phase) from the radius obtained by
integrating the radial velocity measured from spectral lines. This
discrepancy is mostly taken into account by the $p$-factor, but there
may be unaccounted effects introducing an extra phase lag. Finally, even
though the period may be stable for better than $10^{-4}$ days per
cycle, there may still be a variable phase shift between maximum light
and minimum radius, causing a drift of the $\Delta R(\phi_V)/D$ curve
with respect to the lightcurve phase $\phi_L$. This is a possibility
that can potentially affect the phase dependent quantities involved
into the calculations using the BW method, among which the projection
factor $p(\phi)$ and 
the LD correction derived from hydrodynamic models, which require the
exact knowledge of the phase relation between models (computed in
terms of the dynamical phase) and observations (dependent on the
lightcurve phase). There are no indications, however, that such
effect is present. Our best fit results in Table~1, in fact,
suggest that the uncertainty in the phase shift only affects the best
fit diameter for less than 0.001~mas,which is less than 1\% of the
estimated amplitude of the $\zeta$~Gem pulsation. This error is small
enough to be ignored when applying the BW method, compared to the
other larger uncertainties in the data. 

These considerations suggest that, given the current accuracy of
interferometric data, we are not in the position to provide an
independent measurement of the phase shift with better quality than
the available data from \citet{bersier1994a, bersier1994b}. The effect
measured with the independent fit is very small and does not affect
significantly the fit of $R_0$ and $D$. For these reasons we conclude that
a two-parameter fit for the BW method is currently preferable than
leaving the phase shift as a third free parameter.

However, we note that possible variations in the
maximum luminosity -- minimum radius phase shift should be taken into
consideration when the precision of interferometric measurements, and
the accumulation of good quality data over long period of time,
make their direct observation feasible. Even though unpredicted phase
shifts may appear unlikely with our current data, their possible
occurrence should be monitored, as they can provide important insights
on the mechanics of pulsations, and their relation for the stellar
evolution.


\subsection{Limb Darkening}\label{ssec-LD}

The remaining question concerning the accuracy of the BW fit involves
the limb darkening. The LD correction is necessary in order to take into
account the non uniform brightness of the stellar disk, resulting from
the different depths in the stellar atmosphere probed by different
lines of sight. When a star is partially resolved by interferometric
observations, the existence of LD induces a change in the
measured visibility with respect to the simpler case of a uniform
brightness disk. The translation of interferometric visibilities into
angular diameters takes into account this effect, by
fitting the data with limb darkened model visibilities. Alternatively,
when the original data have already been fitted with a UD model, a LD
correction can be applied.

As in the case of the $p$-factor, the LD corrections can be obtained
by solving the dynamic structure of the pulsating stellar atmosphere,
and then computing a complete radiative transfer model to derive the
center-to-limb intensity profile. Using this approach we have derived
a method to compute LD profiles for pulsating Cepheids, presented in
Paper~I. We have then computed specific phase- and
wavelength-dependent LD corrections for $\zeta$~Gem, taking into
account second-order accurate one-dimensional hydrodynamic
calculations performed in spherical geometry, and a full set of atomic
and molecular opacities (\citealt{marengo2003}, hereafter Paper~II).

To convert the UD diameters $\theta_i^{(UD)}$ used in the previous
section into the LD ones, one has to divide by the LD correction
$k(\lambda,\phi)$: 

\begin{equation}\label{eq-LDcorr}
\theta_i^{(LD)} = \frac{\theta_i^{(UD)}}{k(\lambda,\phi)}
\end{equation}

Figure~\ref{fig3} shows the LD corrections we presented in Paper~II
for $\zeta$~Gem. The top curve shows $k$ as a function of the
pulsational phase for the PTI H-band. The model LD correction does
indeed change with phase, with a $\pm 0.3$\% variation around the
average value $\bar k = 0.979$. The largest change occurs close to
minimum radius, when a shock wave crosses the Cepheid atmosphere (see
Paper~I). Barring systematic errors in our models,
we can assume the uncertainty in the phase dependence as the
total error on our LD correction estimate. From Paper~II we have that
this uncertainty results in an extra error of $\sigma_i^{(LD)} \simeq
0.003$~mas ($\sim 3$\% of the $\zeta$~Gem pulsational amplitude),
which is small with respect to the interferometric data errors.
The BW distance
computed with this correction is shown in Figure~\ref{fig4} and in
Table~2. For comparison we have also solved the $\zeta$~Gem BW fit
using the LD correction $k \simeq 0.96 \pm 0.01$ (dashed line in
Figure~\ref{fig3}), derived from
tabulated values computed by \citet{claret1995} for static yellow
supergiants. This values have been used by \citet{lane2002} to derive
the PTI BW distance for $\zeta$~Gem published in their paper.
Given the quoted uncertainty, this results in an extra error source
of $\sigma_i^{(LD)} \simeq 0.02$~mas (as much as 20\% of the
pulsational amplitude).

Figure~\ref{fig4} shows that, within the quoted error uncertainties,
the two solutions for the BW fit with different LD correction are
mutually exclusive, as the respective 1$\sigma$ error regions do not
intersect. In fact, the two LD solutions are also separate from the BW
distance computed for UD diameters. This shows that, even if the final
error range in the best fit $R_0$ and $D$ includes all three
solutions, interferometric techniques can potentially be used to
discriminate between different models of limb darkening. 
Alternatively,  this result can be seen as an indication of how an
independent determination of the LD will result in a much better
measurement of pulsating star distance with the BW method.


\section{Discussion and conclusions}\label{sec-concl}

Table~3 summarizes the error balance of the geometric BW method. For
each error source, the table provides the resulting uncertainty in the
determination of the Cepheid angular diameter, in  mas and as a
fraction of the $\zeta$~Gem 0.1~mas pulsation amplitude. The last
column of the table shows how these uncertainties affect the final
determination of the BW distance. The statistical error in the
interferometric data is still the main source of error in the method,
leading to the final $1 \sigma$ error bars of $\sim 14$\% in our
fit. The systematic errors (related to the radial velocity and phase
shift measurements, the $p$-factor and the limb darkening) still play a
secondary role. Note that the effect of the relative uncertainty of
the systematic error sources, which we have estimated for the
\emph{amplitude} of the stellar pulsation, should be scaled down by
roughly one order of magnitude to estimate their effect on the best
fit distance, since the BW distance depends on the stellar
\emph{radius} which is more than 10 times larger than the pulsation
amplitude. 

The second error source in importance is the determination of the
average value of the LD correction. The difference between the best
fit value of the BW distance obtained with our $\zeta$~Gem model and
with the LD coefficient derived by \citet{claret1995} is $\sim
3$\%. Our $\zeta$~Gem fit shows that using a LD correction that is not
appropriate for the observed Cepheid can introduce an error as
large as when not using LD at all (e.g. using a UD model). This stresses
the importance of reliable modeling of the Cepheid atmosphere even
when using interferometric data with the currently best available
quality. 

The table also shows that the uncertainties in the radial velocity,
phase shift, p-factor and phase dependence of the LD correction at near-IR
wavelength, all together are responsible for less than a 1\% error in
the BW distance determination. These uncertainties plays only a
secondary role compared to the measurement errors. However, with the
constant progress in the development of the interferometric
techniques, the visibility errors are bound to be reduced, and at that
point these other factors, as discussed above, will become
essential to fulfill the potential of the BW method.
This will be especially true for interferometers operating at visible
wavelengths. Figure~\ref{fig3} shows the LD correction we
have computed for $\zeta$~Gem with our model in the V-band (bottom
solid line). The plot
suggests that at certain phases in which the hydrodynamic effects play
an important role (close to minimum radius and in presence of
shockwaves) the interferometric measurement of the Cepheid pulsation
amplitude can be affected by as much as 20\%. This
may result in a further uncertainty of up to 2\% in the BW distance
determination, which can only be corrected with an accurate
time-dependent hydrodynamic modeling of the Cepheid atmosphere. 

Even considering all the error sources discussed in the previous
sections, the current error bar for the $\zeta$~Gem BW diameter is
already two times better than the quoted error in parallax
measurements for this star, which is $\sim$30\% (Hipparcos,
\citealt{esa1997}). A careful analysis of the error regions in the best
fit plane, shows that the error in the best fit angular
diameter is in fact much smaller than the individual errors in $R_0$
and $D$, leaving room for dramatic improvements in the distance
measurement if independent constraints are set on the stellar radius.
These can be derived by modeling the structure of the pulsating
atmosphere. 

The results presented in this work show that a better
determination of the limb darkening of the pulsating star can already
lead to different estimates of the distance and average radius which
are mutually exclusive. This justifies the recent efforts in
producing accurate predictions for the model dependent quantities
needed by the BW method, by means of detailed self-consistent
hydrodynamic models of the pulsating atmosphere (see
e.g. Paper~I and II). An independent determination of the LD
correction for nearby pulsating stars, which will be attainable once
interferometers with baselines of several hundred meters 
will become operative, will thus provide a direct test for such
models. This will open the road for a large scale application of the
geometric BW method, to derive the distances of a large sample of
pulsating stars, and thus attain the goal of an accurate calibration
of their PL relation.


\acknowledgements
We wish to thank the anonymous referee for his/her
comments that helped us to improve this paper. This work was partially
supported by NSF grant AST 98-76734. M.K. is a member of the Chandra
Science Center, which is operated under contract NAS8-39073, and is
partially supported by NASA.


\clearpage



\begin{table}
\begin{center}
\begin{tabular}{lcc}
\multicolumn{3}{c}{TABLE 1}\\
\multicolumn{3}{c}{BW BEST FIT OF $\zeta$~GEM PTI H-BAND UD DATA}\\
\hline
\hline
                  & fixed $\Delta \phi$    & free $\Delta \phi$ \\
\hline
$\Delta \phi$     & 0.280                  & 0.294 \\
$R_0$ [R$_\odot$] & 66.1$^{+8.6}_{-6.8}$   & 66.2$^{+8.6}_{-6.8}$ \\
$D$ [pc]          & 375$^{+49}_{-39}$      & 376$^{+49}_{-39}$ \\
$\Theta_0$ [mas]  & 1.630$\pm 0.007$       & 1.629$\pm 0.007$ \\
\hline
\end{tabular}
\end{center}
\end{table}

\clearpage



\begin{table}
\begin{center}
\begin{tabular}{lcc}
\multicolumn{3}{c}{TABLE 2}\\
\multicolumn{3}{c}{BW BEST FIT OF $\zeta$~GEM PTI H-BAND LD DATA}\\
\hline
\hline
                  & $k = 0.979$            & $k = 0.96$ \\
\hline
$R_0$ [R$_\odot$] & 67.0$^{+8.7}_{-6.9}$   & 66.5$^{+8.5}_{-6.7}$ \\
$D$ [pc]          & 372$^{+49}_{-39}$      & 362$^{+47}_{-37}$ \\
$\Theta_0$ [mas]  & 1.665$\pm 0.007$       & 1.699$\pm 0.007$ \\
\hline
\end{tabular}
\end{center}
\end{table}

\clearpage



\begin{table}
\begin{center}
\begin{tabular}{lccc}
\multicolumn{4}{c}{TABLE 3}\\
\multicolumn{4}{c}{INDIVIDUAL ERROR CONTRIBUTIONS SUMMARY}\\
\hline
\hline
Error source                 & $\sigma$ [mas] & \% of $\zeta$~Gem ampl.
                                              & \% of BW dist. \\
\hline
$\theta_i$ measurement       & 0.01 -- 0.06   & 10 -- 60\%
                                              & 14\% \\
Radial velocity              & 0.002          & 2\%
                                              & 0.2\% \\
p-fact                       & 0.004          & 4\%
                                              & 0.4\% \\
Phase shift                  & $< 0.001$      & $< 1$\%
                                              & $< 0.1$\% \\
LD average value             & 0.02           & 20\%
                                              & 2\% \\
LD phase variations (H band) & $< 0.003$      & $< 3$\%
                                              & $< 0.3$\% \\
LD phase variations (B band) & $<0.02$        & $< 20$\%
                                              & $< 2$\% \\
\hline
\end{tabular}
\end{center}
\end{table}

\clearpage


\figcaption[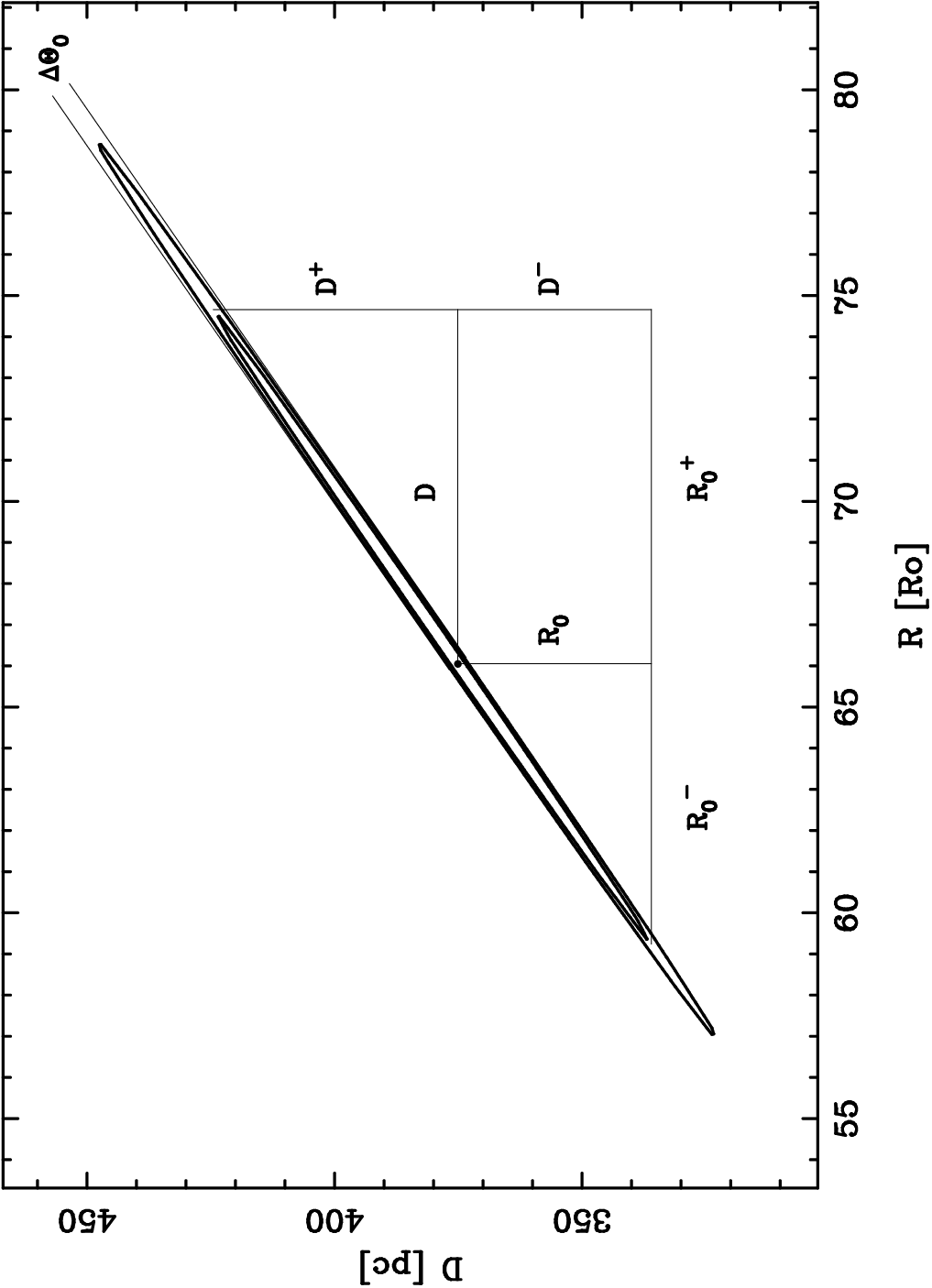]{BW fit of $\zeta$~Gem distance and mean radius
from PTI H-band UD data. Inner contour is 68\% confidence level
(1$\sigma$), and outer contour is 90\% confidence level
error.\label{fig1}}

\figcaption[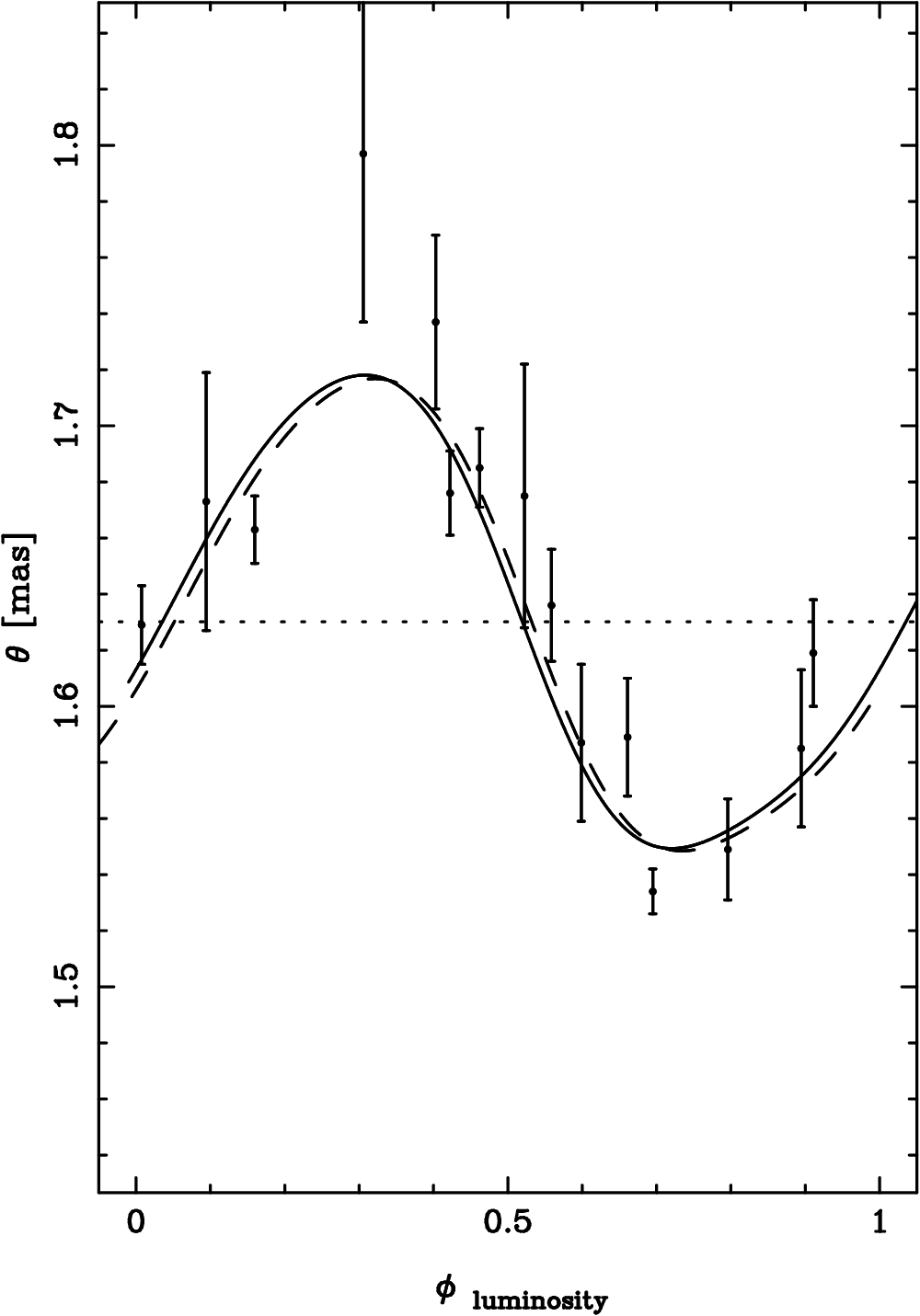]{Best fit UD diameter of $\zeta$~Gem PTI H-band
data from \citet{lane2002}. The solid line is the fixed phase shift as
determined by \citet{bersier1994a, bersier1994b}. The dashed line is
instead the best fit with the free phase shift.\label{fig2}}

\figcaption[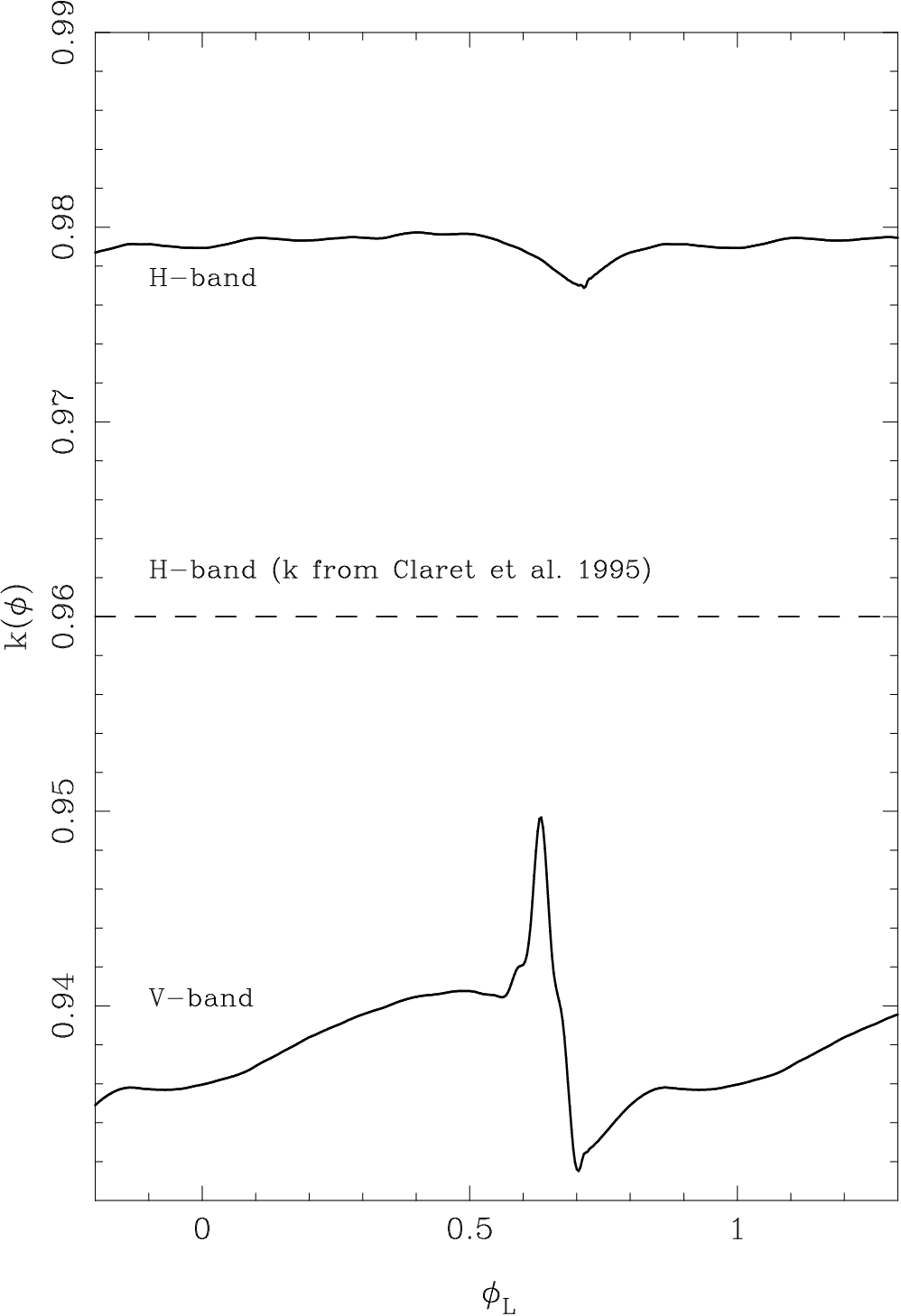]{Phase dependent LD corrections for $\zeta$~Gem
computed by the model presented in Paper~II. The LD correction is
shown for the PTI H-band in the near-IR, and for the optical
V-band.\label{fig3}}

\figcaption[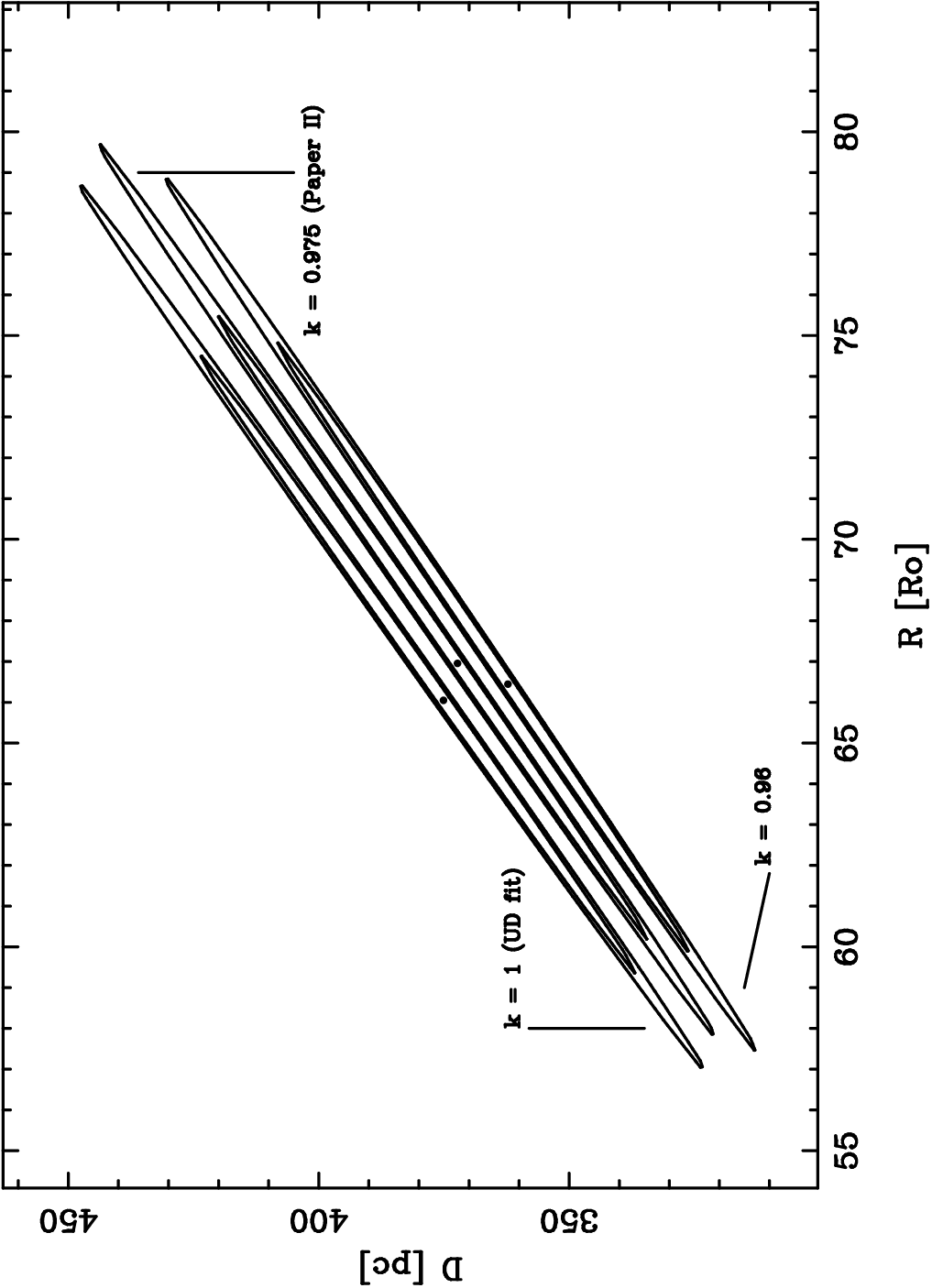]{Best fit parameters and error regions for
$\zeta$~Gem PTI H-band data. Top curve is the UD data, middle curve is
our best fit for model LD data and the bottom curve is the result
obtained using a fixed LD correction of $k \simeq 0.96$ as in
\citet{lane2002}. The inner region is the 68\% confidence level of the
fit (1$\sigma$), while the outer is the 90\% confidence
level.\label{fig4}}


\epsscale{0.85} \plotone{f1.eps} \clearpage
\epsscale{0.85} \plotone{f2.eps} \clearpage
\epsscale{0.85} \plotone{f3.eps} \clearpage
\epsscale{0.85} \plotone{f4.eps} \clearpage


\end{document}